\documentclass[amssymb,11pt]{article}
\usepackage{amsmath}
\usepackage{amsmath}
\usepackage{amssymb}
\usepackage{graphicx}

\def\>{\ensuremath{\rangle}}
\def\<{\ensuremath{\langle}}

\newtheorem{thm}{Theorem}[section]

\newtheorem{lem}{Lemma}[section]
\newtheorem{defn}{Definition}[section]
\newtheorem{prop}{Proposition}[section]
\newtheorem{exam}{Example}[section]

\usepackage[ruled]{algorithm2e}

\SetAlFnt{\small}
\SetAlCapFnt{\small}
\SetAlCapNameFnt{\small}
\SetAlCapHSkip{0pt}
\IncMargin{-\parindent}

\usepackage{times}
\usepackage{amsmath}
\usepackage{stmaryrd}
\usepackage{amssymb}

\def\>{\ensuremath{\rangle}}
\def\<{\ensuremath{\langle}}

\begin{document}
\title{Defining Quantum Control Flow}
\author{Mingsheng Ying, Nengkun Yu and Yuan Feng\\
\\
\small \em QCIS, FEIT, University of Technology, Sydney, 
Australia\\
\small \em and\\
\small \em TNList, Dept. of CS, Tsinghua University, China\\
\small \em Email: Mingsheng.Ying@uts.edu.au, yingmsh@tsinghua.edu.cn}
\date{}
\maketitle

\begin{abstract}
A remarkable difference between quantum and classical programs is that the control flow of the former can be either classical or quantum. One of the key issues in the theory of quantum programming languages is defining and understanding quantum control flow. A functional language with quantum control flow was defined by Altenkirch and Grattage [\textit{Proc. LICS'05}, pp. 249-258]. This paper extends their work, and we introduce a general quantum control structure by defining three new quantum program constructs, namely quantum guarded command, quantum choice and quantum recursion. We clarify  the relation between quantum choices and probabilistic choices. An interesting difference between quantum recursions with classical control flows and with quantum control flows is revealed.\end{abstract}\par
 
\section{Introduction}\label{Intro}

Since Knill~\cite{Kn96} introduced the Quantum Random Access Machine (QRAM) model for quantum computing and proposed a set of conventions for writing quantum pseud-ocodes in 1996, several  quantum programming languages have been defined in the last 16 years; for example QCL by \"{O}mer~\cite{O03}, qGCL by Sanders and Zuliani~\cite{SZ00}, QPL by Selinger~\cite{Se04}, and see~\cite{Ga06} for an excellent survey. One of the key design ideas of almost all existing quantum languages can be summarised by the influential slogan \textquotedblleft quantum data, classical control" proposed by Selinger~\cite{Se04}, meaning that the control flow of a quantum program is still classical, but the program operates on quantum data. An exception is Altenkirch and Grattage's functional language QML~\cite{AG05}, where \textquotedblleft quantum control" flow was introduced; more precisely, they observed that in the quantum setting the case construct naturally splits into two variants: \begin{itemize} 
\item $\mathbf{case}$, which measures a qubit in the data it analyses; \item $\mathbf{case}^\circ$, which analyses quantum data without measuring.
\end{itemize} The control flow in the $\mathbf{case}$ construct is determined by the outcome of a measurement and thus is classical. However, a quantum control flow appears in the $\mathbf{case}^\circ$ construct, as shown in the following example where a special form of the $\mathbf{case}^\circ$, namely the $\mathbf{if}^\circ -\mathbf{then}-\mathbf{else}$ statement, is used.

\begin{exam}\label{qif}Basic quantum gates implemented in QML~\cite{AG05}: The Hadamard gate is written as: \begin{equation*}\begin{split}&had: \mathbf{Q}_2\multimap\mathbf{Q}_2\\
&had\ x=\ \mathbf{if}^\circ\ x\\
&\ \ \ \ \ \ \ \ \ \ \ \ \ \ \ \ \mathbf{then}\ \{\frac{1}{\sqrt{2}}(\mathtt{qfalse}-\mathtt{qtrue})\}\\ 
&\ \ \ \ \ \ \ \ \ \ \ \ \ \ \ \ \mathbf{else}\ \ \ \{\frac{1}{\sqrt{2}}(\mathtt{qfalse}+\mathtt{qtrue})\}\\ 
\end{split}
\end{equation*} and the CNOT gate is as follows: 
\begin{equation*}\begin{split}&cnot: \mathbf{Q}_2\multimap\mathbf{Q}_2\multimap\mathbf{Q}_2\otimes\mathbf{Q}_2\\
&cnot\ c\ x=\ \mathbf{if}^\circ\ c\\
&\ \ \ \ \ \ \ \ \ \ \ \ \ \ \ \ \ \ \ \ \mathbf{then}\ (\mathtt{qtrue}, not\ x)\\ 
&\ \ \ \ \ \ \ \ \ \ \ \ \ \ \ \ \ \ \ \ \mathbf{else}\ \ \ (\mathtt{qfalse}, x)\\ 
\end{split}
\end{equation*} where $\mathbf{Q}_2$ is the type of qubits, and $not$ is the NOT gate: 
\begin{equation*}\begin{split}&not: \mathbf{Q}_2\multimap\mathbf{Q}_2\\
&not\ x=\ \mathbf{if}^\circ\ x\\
&\ \ \ \ \ \ \ \ \ \ \ \ \ \ \ \ \mathbf{then}\ \mathtt{qfalse}\\ 
&\ \ \ \ \ \ \ \ \ \ \ \ \ \ \ \ \mathbf{else}\ \ \ \mathtt{qtrue}\\ 
\end{split}
\end{equation*} 
\end{exam}

A new research line of quantum programming with quantum control flow was then initiated by Altenkirch and Grattage in~\cite{AG05} and further pursued by themselves and others in a series of papers~\cite{AG05a, La08}.

The present paper continues this line of research, and we extend the idea of \textquotedblleft quantum control" by introducing three new quantum program constructs:

\textbf{(1) Quantum Guarded Command:} Our first step toward a general quantum control structure is to introduce a quantum generalisation of Dijkstra's guarded command~\cite{Di75}. Recall that a guarded command can be written as follows: \begin{equation}\label{or-gc}\square_{i=1}^n\ b_i\rightarrow C_i\end{equation} where for each $1\leq i\leq n$, the command $C_i$ is guarded by the Boolean expression $b_i$, and $C_i$ will be executed only when $b_i$ is true. Obviously, the $\mathbf{case}$ operator in QML is a quantum generalisation of guarded command with classical control. On the other hand, as shown in the above example, the $\mathbf{case}^\circ$ operator in QML implements a unitary transformation by decomposing it into two orthogonal branches along the quantum control flow determined by a chosen qubit. So, it is already a kind of guarded command with quantum control flow. 

An even clearer idea for defining quantum guarded command stems from a quite different area, namely quantum walks~\cite{Am01}, \cite{Ah01}:

\begin{exam}\label{qwk} Quantum walks on graphs~\cite{Ah01}: Let $(V,E)$ be an $n-$regular directed graph. Then we can label each edge with a number between $1$ and $n$ such that for each $1\leq i\leq n$, the directed edges labeled $i$ form a permutation. Let $\mathcal{H}_V$ be the Hilbert space spanned by states $\{|v\rangle\}_{v\in V}$. Then for each $1\leq i\leq n$, we can define a shift operator $S_i$ on $\mathcal{H}_V$: $$S_i|v\rangle=|{\rm the}\ i{\rm th\ neighbour\ of}\ v\rangle$$ for any $v\in V$. We introduce an auxiliary quantum variable $q$ with the state Hilbert space $\mathcal{H}_q$ spanned by $\{|i\rangle\}_{i=1}^n$. Now we are able to combine the operators $S_i$ $(1\leq i\leq n)$ along $q$ to form a whole shift operator: \begin{equation}\label{qwk1}S\stackrel{\triangle}{=}\square_{i=1}^n\ q, |i\rangle\rightarrow S_i\end{equation} on $\mathcal{H}_q\otimes\mathcal{H}_V$: \begin{equation}\label{sht}S|v, i\rangle= (S_i|v\rangle)|i\rangle\end{equation} for any $1\leq i\leq n$ and $v\in V$.   
If we further choose a unitary operator $U$ on $\mathcal{H}_q$ then a coined quantum walk on graph $(V,E)$ is defined by modelling its single step by the unitary operator: $$W\stackrel{\triangle}{=}S(I_{\mathcal{H}_V}\otimes U)$$ on $\mathcal{H}_V\otimes\mathcal{H}_q$, where $I_{\mathcal{H}_V}$ is the identity operator in $\mathcal{H}_V$. Usually, $\mathcal{H}_q$ is called the \textquotedblleft coin space", and $U$ the \textquotedblleft coin-tossing operator". 
\end{exam}

The guarded command notation is adopted in Eq.~(\ref{qwk1}) to indicate that the shift operator $S$ is indeed a guarded command with quantum control. It is interesting to note that both Examples~\ref{qif} and \ref{qwk} defined a guarded command with quantum control, but their defining strategies are quite different: in Example~\ref{qif}, a quantum control flow is \textit{detected} by decomposing a unitary operator along an \textit{existing} qubit; in contrast, a quantum control flow is \textit{created} in Example~\ref{qwk} by introducing a \textit{new} quantum variable so that we can combining a family of unitary operators along the created flow. 
The defining strategy used in Example~\ref{qwk} naturally leads us to a general form of quantum guarded command:  
\begin{equation}\label{g-qgc}\square_{i=1}^n\ \overline{q}, |i\rangle\rightarrow P_i\end{equation}
where $P_1,...,P_n$ are a family of quantum programs, and a new family of quantum variables $\overline{q}$ that do not appear in $P_1,...,P_n$ is introduced so that we can form a quantum guarded command by combining $P_1,...,P_n$ along an orthonormal basis $\{|i\rangle\}$ of the state space of $\overline{q}$. For each $1\leq i\leq n$, $P_i$ is guarded by the basis state $|i\rangle$, and a superposition of these basis states yields a quantum control flow. 

\textbf{(2) Quantum Choice:} Guarded commands are the most widely accepted mechanism for nondeterministic programming. Nondeterminism in guarded command~(\ref{or-gc}) is a consequence of the \textquotedblleft overlapping" of the guards $b_1,...,b_n$. In particular, if $b_1=\cdots=b_n=\mathbf{true}$, then guarded command~(\ref{or-gc}) becomes a demonic choice: \begin{equation}\label{dech}\square_{i=1}^n\ C_i,\end{equation} where the alternatives $C_i$ are chosen unpredictably. Usually, the demonic choice is separately defined as an explicit operator rather than a special case of guarded command due to its importance as a means of abstraction in programming. 

To formalise randomised algorithms, research on probabilistic programming \cite{MM05} started in 1980's with the introduction of probabilistic choice:\begin{equation}\label{proch}\square_{i=1}^n\ C_i@p_i,\end{equation} where $\{p_i\}$ is a probability distribution; that is, $p_i\geq 0$ for all $i$, and $\sum_{i=1}^n p_i=1$. The probabilistic choice~(\ref{proch}) randomly chooses the command $C_i$ with probability $p_i$ for every $i$, and thus it can be seen as a refinement of the demonic choice (\ref{dech}). A probabilistic choice is often used to represent a decision in forks according to a certain probability distribution in a randomised algorithm. 

A natural question then arises in the realm of quantum programming: is it possible to define a quantum choice of programs? Indeed, an idea is already there in the construction of quantum walks, although not explicitly stated. In Example~\ref{qwk}, each shift operator $S_i$ can be considered as an independent program, the \textquotedblleft coin-tossing operator" $U$ is employed to create a superposition of $S_i$ $(1\leq i\leq n)$, and thus the single step operator $W$ can be seen as a quantum choice among $S_i$ $(1\leq i\leq n)$. Extending the idea used in Example~\ref{qwk}, we can define a general quantum choice as a sequential composition of a \textquotedblleft coin-tossing" program and a quantum guarded command: \begin{equation}\label{quch}
\square_{i=1}^n\ P;\overline{q},|i\rangle\rightarrow P_i\stackrel{\triangle}{=}P;\square_{i=1}^n\ \overline{q},|i\rangle\rightarrow P_i, 
\end{equation} where $P_1,...,P_n$ are a family of quantum programs, $\overline{q}$ is a new family of quantum variables with $\{|i\rangle\}$ as an orthonormal basis of its state space, and $P$ is a quantum program acting on $\overline{q}$. Intuitively, quantum choice (\ref{quch}) first runs program $P$ to produce a superposition of the execution paths of programs $P_i$ $(1\leq i\leq n)$, and then the guarded command $\square_{i=1}^n\ \overline{q},|i\rangle\rightarrow P_i$ follows. During the execution of the guarded command, each $P_i$ is running along its own path within the whole superposition of execution paths of $P_i$ $(1\leq i\leq n)$. It is widely accepted that quantum superposition is responsible for the advantage of quantum computers over classical computers. The power of superposition of quantum states has been successfully exploited in quantum computing. Quantum choices may provide a platform for explore a higher level of quantum superposition in computing, namely the superposition of quantum programs.

\textbf{(3) Quantum Recursion:} Most classical programming languages allow direct specification of recursive procedures. Quantum loops and more general quantum recursive procedures were already defined in Selinger's language QPL~\cite{Se04}, and termination of quantum loops were analysed by the authors in~\cite{YF10}. But quantum recursions considered in~\cite{Se04, YF10} contain no quantum control flows because there branchings in quantum programs are all determined by the outcomes of quantum measurements. After introducing quantum guarded commands and quantum choices, loops and recursive procedures with quantum control flows can be defined. As will be seen later, a major difference between quantum recursions with and without quantum controls is: auxiliary quantum variables must be introduced in order to define quantum controls. Thus, localisation mechanism is needed in defining quantum recursions with quantum control so that consistency of quantum variables is guaranteed.    

\subsection{Technical Contributions of the Paper}

As shown above, a general notion of quantum control flow comes naturally out from generalising the $\mathbf{case}^\circ$ construct in Altenkirch and Grattage's language QML and the shift operators in quantum walks. However, a major difficulty arises in defining the semantics of quantum guarded commands. For the case where no quantum measurement occur in any $P_i$ $(1\leq i\leq n)$, 
the semantics of each $P_i$ is simply a sequence of unitary operators, and the semantics of guarded command~(\ref{g-qgc}) can be defined in exactly the same way as Eq.~(\ref{sht}). Whenever some $P_i$ contains quantum measurements, however, its semantic structure becomes a tree of linear operators with branching happening at the points where measurements are performed. Then defining the semantics of guarded command (\ref{g-qgc}) requires to properly combine a collection of trees such that certain quantum mechanical principles are obeyed. This problem will be circumvented in Sec.~\ref{sec-comp}.  

\subsection{Organisation of the Paper}

A new quantum programming language QGCL with quantum guarded commands is defined in Sec. \ref{syntax}. Sec. ~\ref{sec-comp} prepares some key ingredients needed in defining the semantics of QGCL. The denotational semantics and weakest precondition semantics of QGCL are presented in Sec.~\ref{Seman}. In Sec.~\ref{QChoice}, quantum choice is defined in terms of quantum guarded command, and probabilistic choice is implemented by quantum choice by introducing local variables. Because of the limited space, quantum recursion is only briefly touched in Sec.~\ref{QREC}. For readability, all proofs are deferred to the Appendix.  

\section{QGCL: A Language with Quantum Guarded Commands}\label{syntax}

We now define a quantum programming language QGCL with quantum guarded commands. QGCL is essentially an extension of Sanders and Zuliani's qGCL obtained by adding quantum control flow. But the presentation of QCGL is quite different from qGCL due to the complications in the semantics of quantum guarded commands.
We assume a countable set $qVar$ of quantum variables ranged over by $q,q_1,q_2,...$. For simplicity of the presentation, we only consider a purely quantum programming language, but we include a countably infinite set $Var$ of classical variables ranged over by $x,y,...$ so that we can use them to record outcomes of quantum measurements. However, classical computation described by, for example, the assignment statement $x:=e$ in a classical programming language is excluded.      
It is required that the sets of classical and quantum variables are disjoint. For each classical variable $x\in Var$, its type is assumed to be a non-empty set $D_x$; that is, $x$ takes values from $D_x$. For each quantum variable $q\in qVar$, its type is a Hilbert space $type(q)=\mathcal{H}_q$, which is the state space of the quantum system denoted by $q$. For a sequence $\overline{q}=q_1,q_2,\cdots$ of quantum variables, we write: $$type(\overline{q})=\mathcal{H}_{\overline{q}}=\bigotimes_{i\geq 1}\mathcal{H}_{q_i}.$$ Similarly, for any set $V\subseteq qVar$, we write: $$type(V)=\mathcal{H}_V=\bigotimes_{q\in V}\mathcal{H}_q.$$ In particular, we write $\mathcal{H}_{all}$  for $type(qVar).$ To simplify the notation, we often identify a sequence of variables with the set of these variables provided they are distinct.    

\begin{defn}\label{syn-def} For each QGCL program $P$, we write $var(P)$ for the set of its classical variables and $qvar(P)$ for its quantum variables. QGCL programs are inductively defined as follows:\begin{enumerate}
\item $\mathbf{abort}$ and $\mathbf{skip}$ are programs, and $$var(\mathbf{abort})=var(\mathbf{skip})=\emptyset,$$ $$qvar(\mathbf{abort})=qvar(\mathbf{skip})=\emptyset.$$
\item If $\overline{q}=q_1,...,q_k$ is a sequence of quantum variables, and $U$ is a unitary operator on $type(\overline{q})$, then $U[\overline{q}]$ is a program, and
$$var(U[\overline{q}])=\emptyset,\ \ \ qvar(U[\overline{q}])=\overline{q}.$$
\item If $\overline{q}=q_1,...,q_k$ is a sequence of quantum variables, $x$ is a classical variable, $M=\{M_m\}$ is a quantum measurement in $type(\overline{q})$, and $\{P_m\}$ is a family of programs indexed by the outcomes $m$ of measurement $M$ such that $spec(M)\subseteq D_x,$ where $spec(M)=\{m\}$ is the spectrum of $M$; that is, the set of all possible outcomes of $M$, and $x\notin \bigcup_{m}var(P_m),$ then 
\begin{equation}\label{me-def}P\stackrel{\triangle}{=}M[x\leftarrow\overline{q}]:\{P_m\}\end{equation} is a program, and $$var(P)=\{x\}\cup\bigcup_mvar(P_m),$$ $$qvar(P)=\overline{q}\cup\bigcup_m qvar(P_m).$$
\item If $\overline{q}=q_1,...,q_k$ is a sequence of quantum variables, $\{|i\rangle\}_{i=1}^n$ is an orthonormal basis of $type(\overline{q})$, and $\{P_i\}_{i=1}^n$ is a family of programs such that 
$$\overline{q}\cap\bigcup_{i=1}^n qVar(P_i)=\emptyset,$$ then $$P\stackrel{\triangle}{=}\square_{i=1}^n\ \overline{q},\ |i\rangle\rightarrow P_i$$ is a program, and 
$$var(P)=\bigcup_{i=1}^n var(P_i),$$ $$qvar(P)=\overline{q}\cup\bigcup_{i=1}^n qvar(P_i).$$\item If $P_1$ and $P_2$ are programs such that $var(P_1)\cap var(P_2)=\emptyset$, then $P_1;P_2$ is a program, and $$var(P_1;P_2)=var(P_1)\cup var(P_2),$$ $$qvar(P_1;P_2)=qvar(P_1)\cup qvar(P_2).$$
\end{enumerate}\end{defn}

The meanings of $\mathbf{abort}$ and $\mathbf{skip}$ are the same as in a classical programming language. Two kinds of statements are introduced in the above definition to describe basic quantum operations, namely unitary transformation and measurement. In the unitary transformation $U[\overline{q}]$, only quantum variables $\overline{q}$ but no classical variables appear, and the transformation is applied to $\overline{q}$. In statement (\ref{me-def}), a measurement $M$ is first performed on quantum variables $\overline{q}$ with the outcome stored in classical variable $x$, and then whenever outcome $m$ is reported, the corresponding subprogram $P_m$ is executed. The intuitive meaning of quantum guarded command was already carefully explained in Sec.~\ref{Intro}.   
Whenever the sequence $\overline{q}$ of quantum variables can be recognised from the context, $\square_{i=1}^n\ \overline{q},\ |i\rangle\rightarrow P_i$ can be abbreviated to $\square_{i=1}^n\ |i\rangle\rightarrow P_i.$ The sequential composition $P_1;P_2$ is similar to that in a classical language, and the requirement $var(P_1)\cap var(P_2)=\emptyset$ means that the outcomes of measurements performed at different points are stored in different classical variables. Such a requirement is mainly for technical convenience, and it will considerably simplify the presentation.
The syntax of QGCL can be summarised as follows:
\begin{equation}\label{qudef}\begin{split}P:=\ &\mathbf{abort}\ |\ \mathbf{skip}\ |\ P_1;P_2\\ 
 &|\ U[\overline{q}] \ \ \ \ \ \ \ \ \ \ \ \ \ \ \ \ \ \ \ \ \ \ \ \ ({\rm unitary\ transformation})\\ 
&|\ \mathbf{measure}\ M[\overline{q}]:\{P_m\}\ ({\rm quantum\ measurement}\\ & \ \ \ \ \ \ \ \ \ \ \ \ \ \ \ \ \ \ \ \ \ \ \ ={\rm classical\ guarded\ command}) \\ &|\ \square_{i=1}^n\ \overline{q}, |i\rangle\rightarrow P_i \ \ \ ({\rm quantum\ guarded\ command})
\end{split}\end{equation} 

\section{Guarded Compositions of Quantum Operations}\label{sec-comp}

\subsection{Guarded composition of unitary operators}

A major difficulty in defining the semantics of QGCL comes from the treatment of guarded commands where a guarded composition of semantic functions is vital. To ease the understanding of a general definition of such a guarded composition, we start with the guarded composition of unitary operators, which is a straightforward generalisation of the quantum walk shift operator $S$ in Example~\ref{qwk}. 

\begin{defn}\label{eqU}For each $1\leq i\leq n$, let $U_i$ be an unitary operator in Hilbert space $\mathcal{H}$. Let $\mathcal{H}_s$ be a Hilbert space with $\{|i\rangle\}$ as an orthonormal basis. Then we define a linear operator: $$U\stackrel{\triangle}{=}\square_{i=1}^n\ |i\rangle\rightarrow U_i$$ in $\mathcal{H}\otimes\mathcal{H}_s$ by $$U(|\psi\rangle|i\rangle)=(U_i|\psi\rangle)|i\rangle$$ for any $|\psi\rangle\in\mathcal{H}$ and for any $1\leq i\leq n$. Then by linearity we have: \begin{equation}\label{eqUU}U\left(\sum_{i=1}^n|\psi_i\rangle |i\rangle\right)=\sum_{i=1}^n(U_i|\psi_i\rangle)|i\rangle\end{equation} for any $|\psi_1\rangle,...,|\psi_n\rangle\in\mathcal{H}$. The operator $U$ is called the guarded composition of $U_i$ $(1\leq i\leq n)$ along $\{|i\rangle\}$.\end{defn}

\begin{exam}Quantum multiplexor: As a quantum generalisation of multiplexor, a well-known notion in digit logic, quantum multiplexor (QMUX for short) was introduced in~\cite{SBM06} as a useful tool in synthesis of quantum circuits. A QMUX $U$ with $k$ select qubits and $d-$qubit-wide data bus can be represented by a block-diagonal matrix:   
$$U=diag (U_0,U_1,...,U_{2^k-1})=\left(\begin{array}{cccc}U_0& & & \\
& U_1& & \\ & & ... & \\ & & & U_{2^k-1}
\end{array}\right).$$ Multiplexing $U_0,U_1,...,U_{2^k-1}$ with $k$ select quits is exactly the guarded composition $$\square_{i=0}^{2^k-1}|i\rangle\rightarrow U_i$$ along the computational basis of $k$ qubits. \end{exam}

\begin{lem}The guarded composition $\square_{i=1}^n\ |i\rangle\rightarrow U_i$ is an unitary operator in $\mathcal{H}\otimes\mathcal{H}_s$.
\end{lem}

\subsection{Operator-valued functions}

For any Hilbert space $\mathcal{H}$, we write $\mathcal{L}(\mathcal{H})$ for the space of (linear) operators on $\mathcal{H}$.

\begin{defn} Let $\Delta$ be a nonempty set. Then a function $F:\Delta\rightarrow \mathcal{L}(\mathcal{H})$ is called an operator-valued function in $\mathcal{H}$ over $\Sigma$ if \begin{equation}\label{sem-con}\sum_{\delta\in \Delta}F(\delta)^\dag\cdot F(\delta)\sqsubseteq I_{\mathcal{H}},\end{equation} where $I_\mathcal{H}$ is the identity operator in $\mathcal{H}$, and $\sqsubseteq$ stands for the L\"{o}wner order; that is, $A\sqsubseteq B$ if and only if $B-A$ is a positive operator. In particular, $F$ is said to be full when Eq.~(\ref{sem-con}) becomes equality. 
\end{defn}

The simplest examples of operator-valued function are unitary operators and measurements. 

\begin{exam}\begin{enumerate}\item A unitary operator on Hilbert space $\mathcal{H}$ can be seen as a full operator-valued function over a singleton $\Delta=\{\epsilon\}$.
\item A measurement $M$ on Hilbert space $\mathcal{H}$ can be seen as a full operator-valued function over its spectrum $Spec(M).$
\end{enumerate}\end{exam}

More generally, a super-operator defines a family of operator-valued functions. Let $\mathcal{E}$ be a super-operator on Hilbert space $\mathcal{H}$. Then $\mathcal{E}$ has the Kraus operator-sum representation: $\mathcal{E}=\sum_i E_i\circ E_i^\dag,$ meaning: $\mathcal{E}(\rho)=\sum_i E_i\rho E_i^\dag$ for all density operators $\rho$ in $\mathcal{H}$. For such a representation, we set $\Delta=\{i\}$ for the set of indexes, and define an operator-valued function over $\Delta$ by $F(i)=E_i$ for every $i$. Since operator-sum representation of $\mathcal{E}$ is not unique, $\mathcal{E}$ defines not only a single operator-valued function. We write $\mathbb{F}(\mathcal{E})$ for the family of operator-valued functions defined by all Kraus operator-sum representations of $\mathcal{E}$. Conversely, an operator-valued function determines uniquely a super-operator. 

\begin{defn}Let $F$ be an operator-valued function in Hilbert space $\mathcal{H}$ over set $\Delta$. Then $F$ defines a super-operator $\mathcal{E}(F)$ in $\mathcal{H}$ as follows: $$\mathcal{E}(F)=\sum_{\delta\in\Delta} F(\delta)\circ F(\delta)^\dag.$$
\end{defn}

For a family $F$ of operator-valued functions, we write $\mathcal{E}(\mathbb{F})=\{\mathcal{E}(F):F\in\mathbb{F}\}.$ It is obvious that $\mathcal{E}(\mathbb{F}(\mathcal{E}))=\{\mathcal{E}\}.$ On the other hand, for any operator-valued function $F$ over $\Delta=\{\delta_1,...,\delta_k\}$, Theorem 8.2 in \cite{NC00} indicates that $\mathbb{F}(\mathcal{E}(F))$ consists of all operator-valued functions $G$ over $\Gamma=\{\gamma_1,...,\gamma_l\}$ such that $$G(\gamma_i)=\sum_{j=1}^n u_{ij}\cdot F(\delta_j)$$ for each $1\leq i\leq n$, where $n=\max (k,l)$, $U=(u_{ij})$ is an $n\times n$ unitary matrix, $F(\delta_i)=G(\gamma_j)=0_\mathcal{H}$ for all $k+1< i\leq n$ and $l+1< j\leq n$.
 
\subsection{Guarded composition of operator-valued functions}

We first introduce a notation. Let $\Delta_i$ be a nonempty set for every $1\leq i\leq n$. Then the superposition of $\Delta_i$ $(1\leq i\leq n)$ is defined as follows: 
$$\bigoplus_{i=1}^n\Delta_i=\{\oplus_{i=1}^n\delta_i:\delta_i\in\Delta_i\ {\rm for\ every}\ 1\leq i\leq n\}.$$

\begin{defn}\label{qgu}For each $1\leq i\leq n$, let $F_i$ be an operator-valued function in Hilbert space $\mathcal{H}$ over set $\Delta_i$. Let $\mathcal{H}_s$ be a Hilbert space with $\{|i\rangle\}$ as an orthonormal basis. Then the guarded composition of $F_i$ $(1\leq i\leq n)$ along $\{|i\rangle\}$ is defined to be the operator-valued function in $\mathcal{H}\otimes\mathcal{H}_s$ over $\bigoplus_{i=1}^n\Delta_i$: 
$$F\stackrel{\triangle}{=}\square_{i=1}^n\ |i\rangle\rightarrow F_i,$$  
\begin{equation}\label{coef0}F(\oplus_{i=1}^n\delta_i)\left(\sum_{i=1}^n|\psi_i\rangle|i\rangle\right)=\sum_{i=1}^n\left(\prod_{k\neq i}\lambda_{k\delta_k}\right)(F_i(\delta_i)|\psi_i\rangle)|i\rangle\end{equation} for any $|\psi_1\rangle, ..., |\psi_n\rangle\in\mathcal{H}$ and for any $\delta_i\in\Delta_i$ $(1\leq i\leq n)$, where \begin{equation}\label{coef1}\lambda_{k\delta_k}=\sqrt{\frac{tr F_k(\delta_k)^\dag F_{k}(\delta_k)}{\sum_{\tau_k\in\Delta_k}tr F_{k}(\tau_k)^\dag F_{k}(\tau_k)}}.\end{equation} In particular, if $F_k$ is full and $d=\dim\mathcal{H}<\infty$, then $$\lambda_{k\delta_k}=\sqrt{\frac{tr F_{k}(\delta_k)^\dag F_{k}(\delta_k)}{d}}$$ for any $\delta_k\in\Delta_k$ $(1\leq k\leq n)$.
\end{defn}

It is easy to see that whenever $\Delta_i$ is a singleton for all $1\leq i\leq n$, then Eq.~(\ref{coef0}) degenerates to Eq.~(\ref{eqUU}). So, the above definition is a generalisation of Definition~\ref{eqU}.

\begin{exam}\label{gmeas} (Guarded composition of measurements) We consider two simplest measurements; that is, measurements on a qubit in the computational basis $|0\rangle, |1\rangle$ and in basis $|\pm\rangle=\frac{1}{\sqrt{2}}(|0\rangle\pm |1\rangle)$: 
$$M^0=\{M_0^0=|0\rangle\langle 0|,M_1^0=|1\rangle\langle 1|\},$$ $$M^1=\{M_0^1=|+\rangle\langle +|,M_1^1=|-\rangle\langle -|\}.$$ Then their guarded composition along another qubit is  measurement \begin{equation*}\begin{split}M&=(|0\rangle\rightarrow M_0)\ \square\ (|1\rangle\rightarrow M_1)\\ &=\{M_{00}, M_{01}, M_{10}, M_{11}\}\end{split}\end{equation*} on two qubits, where 
$$M_{ij}(|\psi_0\rangle|0\rangle+|\psi_1\rangle|1\rangle)=\frac{1}{\sqrt{2}}(M_i^0|\psi_0\rangle|0\rangle+M_j^1|\psi_1\rangle|1\rangle)$$ for any states $|\psi_0\rangle, |\psi_1\rangle$ of a qubit and $i,j\in\{0,1\}$.
\end{exam}

The following lemma shows that the guarded composition of operator-valued functions is well-defined.

\begin{lem}\label{gulem}The guarded composition $F\stackrel{\triangle}{=}\square_{i=1}^n\ |i\rangle\rightarrow F_i$ is an operator-valued function in $\mathcal{H}\otimes\mathcal{H}_s$ over $\bigoplus_{i=1}^n\Sigma_i$. In particular, if all $F_i$ $(1\leq i\leq n)$ are full, then so is $F$.   
\end{lem}

\subsection{Guarded composition of super-operators}

Guarded composition of a family of super-operators can be defined through the guarded compsition of the operator-valued functions generated from them. 

\begin{defn}\label{def-gsup}For each $1\leq i\leq n$, let $\mathcal{E}_i$ be a super-operator in Hilbert space $\mathcal{H}$. Let $\mathcal{H}_s$ be a Hilbert space with $\{|i\rangle\}$ as an orthonormal basis. Then the guarded composition of $\mathcal{E}_i$ $(1\leq i\leq n)$ is defined to be the family of super-operators: 
\begin{equation*}\begin{split}\square_{i=1}^n\ |i\rangle\rightarrow\mathcal{E}_i&=\{\mathcal{E}(\square_{i=1}^n\ |i\rangle\rightarrow F_i):\\ & \ \ \ \ \ \ \ \ \ \ \ \ F_i\in\mathbb{F}(\mathcal{E}_i)\ {\rm for\ every}\ 1\leq i\leq n\}.\end{split}\end{equation*}
\end{defn}

It is easy to see that if $n=1$ then the above guarded composition of super-operators consists of only $\mathcal{E}_1$. For $n>1$, however,  it is not a singleton, as shown by the following:

\begin{exam}Let $\mathcal{E}_0$ and $\mathcal{E}_1$ be the super-operators in Hilbert space $\mathcal{H}$ defined by unitary operators $U_0$, $U_1$, respectively; that is, $\mathcal{E}_i=U_i\circ U_i^\dag$ $(i=0,1)$. We set $U$ to be the guarded composition of $U_0$ and $U_1$: $U=(|0\rangle\rightarrow U_0)\square (|1\rangle\rightarrow U_1).$ Then the super-operator defined by $U$ is $\mathcal{E}(U)\in (|0\rangle\rightarrow \mathcal{E}_0)\square (|1\rangle\rightarrow \mathcal{E}_1).$ Indeed, we have: 
$$(|0\rangle\rightarrow \mathcal{E}_0)\square (|1\rangle\rightarrow \mathcal{E}_1)=\{\mathcal{E}_\theta=U_\theta\circ\ U_\theta^\dag:0\leq \theta<2\pi\},$$ where 
$U_\theta=(|0\rangle\rightarrow U_0)\square (|1\rangle\rightarrow e^{i\theta} U_1).$
The non-uniqueness of the members of the above guarded composition is caused by the relative phase $\theta$ between $U_0$ and $U_1$.   
\end{exam}

\section{Semantics of QGCL}\label{Seman}

We first introduce several notations needed in this section. Let $\mathcal{H}$ and $\mathcal{H}^\prime$ be two Hilbert spaces, and let $E$ be an operateor in $\mathcal{H}$. Then the cylindrical extension of $E$ in $\mathcal{H}\otimes\mathcal{H}^\prime$ is defined to be the operator $E\otimes I_{\mathcal{H}^\prime}$, where $I_{\mathcal{H}^\prime}$ is the identity operator in $\mathcal{H}^\prime$. For simplicity, we will write $E$ for $E\otimes I_{\mathcal{H}^\prime}$ whenever confusion does not happen. Let $F$ be an operator-valued function in $\mathcal{H}$ over $\Sigma$. Then the cylindrical extension of $F$ in $\mathcal{H}\otimes\mathcal{H}^\prime$ is the operator-valued function $F^\prime$ in $\mathcal{H}\otimes\mathcal{H}^\prime$ over $\Delta$ defined by $F^\prime(\delta)=F(\delta)\otimes I_{\mathcal{H}^\prime}$ for every $\delta\in\Delta$. For simplicity, we often write $F$ for $F^\prime$ whenever confusion can be excluded from the context. 
Furthermore, let $\mathcal{E}=\sum_i E_i\circ E_i^\dag$ be a super-operator in $\mathcal{H}$. Then the cylindrical extension of $\mathcal{E}$ in $\mathcal{H}\otimes\mathcal{H}^\prime$ is defined to be the super-operator: $\mathcal{E}=\sum_i(E_i\otimes I_{\mathcal{H}^\prime})\circ (E_i^\dag\otimes I_{\mathcal{H}^\prime}).$ Here, for simplicity, the same symbol $\mathcal{E}$ is used to denote the extension of $\mathcal{E}$. In particular, if $E$ is an operator in $\mathcal{H}$, and $\rho$ is a partial density operator in $\mathcal{H}\otimes\mathcal{H}^\prime$, then $E\rho E^\dag$ should be understood as $(E\otimes I_{\mathcal{H}^\prime})\rho(E^\dag\otimes I_{\mathcal{H}^\prime})$. If $\mathcal{E}_1$ and $\mathcal{E}_2$ are two super-operators in a Hilbert space $\mathcal{H}$, then their (sequential) composeition $\mathcal{E}_1;\mathcal{E}_2$ is the super-operator in $\mathcal{H}$ defined by $(\mathcal{E}_1;\mathcal{E}_2)(\rho)=\mathcal{E}_2(\mathcal{E}_1(\rho))$ for any partial density operator $\rho$ in $\mathcal{H}$.  

\subsection{Classical states}

We now define the states of classical variables in QGCL. 

\begin{defn}\label{cl-state}The (partial) classical states and their domains are inductively defined as follows:\begin{enumerate} \item $\epsilon$ is a classical state, called the empty state, and $dom(\epsilon)=\emptyset$; \item If $x\in Var$ is a classical variable, and $a\in D_x$ is an element of the domain of $x$, then $[x\leftarrow a]$ is a classical state, and $dom([x\leftarrow a])=\{x\}$; \item If both $\delta_1$ and $\delta_2$ are classical states, and $dom(\delta_1)\cap dom(\delta_2)=\emptyset$, then $\delta_1\delta_2$ is a classical state, and $dom(\delta_1\delta_2)=dom(\delta_1)\cup dom(\delta_2)$; \item If $\delta_i$ is a classical state for every $1\leq i\leq n$, then $\oplus_{i=1}^n\delta_i$ is a classical state, and $$dom(\oplus_{i=1}^n\delta_i)=\bigcup_{i=1}^n dom(\delta_i).$$  
\end{enumerate}
\end{defn}

Intuitively, a classical state $\delta$ defined by clauses (1) to (3) in the above definition can be seen as a (partial) assignment to classical variables; more precisely, $\delta$ is an element of $\delta\in\prod_{x\in dom(\delta)}D_x;$ that is, a choice function: $\delta:V\rightarrow\bigcup_{x\in dom(\delta)}D_x$ such that $\delta(x)\in D_x$ for every $x\in dom(\delta)$. In particular, $\epsilon$ is the empty function. Since $\prod_{x\in\emptyset}D_x=\{\epsilon\},$ $\epsilon$ is the only possible state of with empty domain. The state $[x\leftarrow a]$ assigns value $a$ to variable $x$ but the values of the other variables are undefined. The composed state $\delta_1\delta_2$ can be seen as the assignment to variables in $dom(\delta_1)\cup dom(\delta_2)$ given by  
\begin{equation}\label{comp-sta}(\delta_1\delta_2)(x)=\begin{cases}\delta_1(x) &{\rm if}\ x\in dom(\delta_1),\\ \delta_2(x) &{\rm if}\ x\in dom(\delta_2).\end{cases}\end{equation}
Eq.~(\ref{comp-sta}) is well-defined since it is required that $dom(\delta_1)\cap dom(\delta_2)=\emptyset.$ In particular, $\epsilon\delta=\delta\epsilon=\delta$ for any state $\delta$, and  
if $x\notin dom(\delta)$ then $\delta [x\leftarrow a]$ is the assignment to variables in $dom(\delta)\cup\{x\}$ given by  
$$\delta[x\leftarrow a](y)=\begin{cases}
\delta(y) &{\rm if}\ y\in dom(\delta),\\ a &{\rm if}\ y=x.
\end{cases}$$ The state $\oplus_{i=1}^n\delta_i$ defined by clause (4) in Definition~\ref{cl-state} can be thought of as a kind of superposition of $\delta_i$ $(1\leq i\leq n)$. 
 
\subsection{Semi-classical denotational semantics}

For each QGCL program $P$, we write $\Delta(P)$ for the set of all possible states of its classical variables. The semi-classical denotational semantics $\lceil P\rceil$ of $P$ will be defined as an operator-valued function in $\mathcal{H}_{qvar(P)}$ over $\Delta(P)$, where $\mathcal{H}_{qvar(P)}$ is the type of quantum variables occurring in $P$. In particular, if $qvar(P)=\emptyset$; for example $P=\mathbf{abort}$ or $\mathbf{skip}$, then $\mathcal{H}_{qvar(P)}$ is a one-dimensional space $\mathcal{H}_\emptyset$, and an operateor in $\mathcal{H}_\emptyset$ can be identified with a complex number; for instance the zero operator is number $0$ and the identity operator is number $1$. For any set $V\subseteq qVar$ of quantum variables, we write $I_V$ for the identity operator in Hilbert space $\mathcal{H}_V$.

\begin{defn}\label{defsem}The classical state $\Delta(P)$ and semi-classical semantic function $\lceil P\rceil$ of a QGCL program $P$ are inductively defined as follows:\begin{enumerate} \item $\Delta(\mathbf{abort})=\{\epsilon\}$, and $\lceil\mathbf{abort}\rceil(\epsilon)=0$; \item $\Delta(\mathbf{skip})=\{\epsilon\}$, and $\lceil\mathbf{skip}\rceil(\epsilon)=1;$ \item $\Delta(U[\overline{q}])=\{\epsilon\}$, and $\lceil U[\overline{q}]\rceil(\epsilon) =U_{\overline{q}},$ where $U_{\overline{q}}$ is the unitary operator $U$ acting in $\mathcal{H}_{\overline{q}}$; 
\item If $P\stackrel{\triangle}{=}M[x\leftarrow\overline{q}]:\{P_m\},$ where $M=\{M_m\}$, then $$\Delta(P)=\bigcup_m\{\delta[x\leftarrow m]:\delta\in D(P_m)\},$$ $$\lceil P\rceil (\delta[x\leftarrow m])=(\lceil P_m\rceil (\delta)\otimes I_{V\setminus qvar(P_m)})\cdot (M_m\otimes I_{V\setminus\overline{q}})$$ for every $\delta\in \Delta(P_m)$ and for every $m$, where $V=\overline{q}\cup\bigcup_{m}qvar(P_m)$; \item If $P\stackrel{\triangle}{=}\square_{i=1}^n\ \overline{q},|i\rangle\rightarrow P_i,$ then $$\Delta(P)=\bigoplus_{i=1}^n\Delta(P_i),$$ $$\lceil P\rceil =\square_{i=1}^n\ |i\rangle\rightarrow \lceil P_i\rceil;$$
\item \begin{equation}\label{sem-seq}\begin{split}\Delta &(P_1;P_2)=\Delta(P_1);\Delta(P_2)\\ &=\{\delta_1\delta_2:\delta_1\in\Delta(P_1)\ {\rm and}\ \delta_2\in\Delta(P_2)\},\end{split}\end{equation}
\begin{equation*}\begin{split}\lceil P_1;P_2\rceil (\delta_1\delta_2)=(\lceil &P_2\rceil (\delta_2)\otimes I_{V\setminus qvar(P_2)})\\ &\cdot (\lceil P_1\rceil (\delta_1)\otimes I_{V\setminus qvar(P_1)})\end{split}\end{equation*} where $V=qvar(P_1)\cup qvar(P_2)$;
\end{enumerate}
\end{defn}

Since it is required in Definition~\ref{syn-def} that $var(P_1)\cap var(P_2)=\emptyset$ in the sequential composition $P_1;P_2$, we have $dom(\delta_1)\cap dom(\delta_2)=\emptyset$ for any $\delta_1\in\Delta(P_1)$ and $\delta_2\in\Delta(P_2)$. Thus, Eq.~(\ref{sem-seq}) is well-defined. Intuitively, if a quantum program $P$ does not contain any guarded command, then its semantic structure can be seen as a tree with its nodes labelled by basic commands and its edges by linear operators. This tree grows up from the root in the following way: if the current node is labelled by a unitary transformation $U$, then a single edge stems from the node and it is labelled by $U$; and if the current node is labelled by a measurement $M=\{M_m\}$, then for each possible outcome $m$, an edge stems from the node and it is labelled by the measurement operator $M_m$. Thus, each classical state $\delta\in\Delta(P)$ is corresponding to a branch in the semantic tree of $P$, and the value of semantic function $\lceil P\rceil$ in state $\delta$ is the (sequential) composition of the operators labelling the edges of $\delta$. The semantic structure of a quantum program $P$ with guarded commands is much more complicated. We can imagine it as a tree with superpositions of nodes that generate superpositions of branches. The value of semantic function $\lceil P\rceil$ in a superpositions of branches is then defined as the guarded composition of the values in these branches.        

\subsection{Purely quantum denotational semantics}

Now the purely quantum semantics of a quantum program can be naturally defined as the super-operator induced by its semi-classical semantic function.

\begin{defn}\label{den-def}For each QGCL program $P$, its purely quantum denotational semantics is the super-operator $\llbracket P\rrbracket$ in $\mathcal{H}_{qvar(P)}$ defined as follows:  
$$\llbracket P\rrbracket =\mathcal{E}(\lceil P\rceil)=\sum_{\delta\in \Delta(P)}\lceil P\rceil (\delta)\circ \lceil P\rceil(\delta)^\dag.$$
\end{defn}

The following proposition presents a representation of the purely quantum semantics of a program in terms of its subprograms. 

\begin{prop}\label{qsem}\begin{enumerate}
\item $\llbracket\mathbf{abort}\rrbracket =0;$
\item $\llbracket\mathbf{skip} \rrbracket =1;$
\item $\llbracket P_1;P_2\rrbracket  =\llbracket P_1\rrbracket ;\llbracket P_2\rrbracket;$
\item $\llbracket U[\overline{q}]\rrbracket = U_{\overline{q}}\circ U_{\overline{q}};$
\item $\llbracket M[x\leftarrow\overline{q}]:\{P_m\}\rrbracket =\sum_m \left[(M_m\circ M_m^\dag);\llbracket P_m\rrbracket\right].$ Here, $\llbracket P_m\rrbracket$ should be seen as a cylindrical extension in $\mathcal{H}_{V}$ from $\mathcal{H}_{qvar(P_m)}$, $M_m\circ M_m^\dag$ as a cylindrical extension in $\mathcal{H}_{V}$ from $\mathcal{H}_{\overline{q}}$, and $V=\overline{q}\cup\bigcup_{m} qvar(P_m)$;  
\item $\llbracket \square_{i=1}^n\ \overline{q}, |i\rangle\rightarrow P_i\rrbracket\in \square_{i=1}^n\ |i\rangle\rightarrow \llbracket P_i\rrbracket.$ Here $\llbracket P_i\rrbracket$ should be understood as a cylindrical extension in $\mathcal{H}_{V}$ from $\mathcal{H}_{qvar(P_i)}$ for every $1\leq i\leq n$, and $V=\overline{q}\cup\bigcup_{i=1}^n qvar(P_i)$.\end{enumerate}\end{prop}

The symbol \textquotedblleft$\in$" in clause 6) of the above proposition can be understood as a refinement relation. It is worth noting that in general \textquotedblleft$\in$" cannot be replaced by equality. This is exactly the reason that the purely quantum semantics of a program has to be derived through its semi-classical semantics and cannot be defined directly in a compositional way.  

Equivalence relation between quantum programs can be introduced based on their purely quantum semantics. 

\begin{defn}Let $P$ and $Q$ be two QGCL programs. If $qvar(P)=qvar(Q)$ and $\llbracket P\rrbracket =\llbracket Q\rrbracket$, then we say that $P$ and $Q$ are equivalent and write $P\equiv Q$. 
\end{defn}

\subsection{Weakest Precondition Semantics}

The notion of quantum weakest precondition was proposed by D'Hondt and Panangaden~\cite{DP06}.  

\begin{defn} Let $P$ be a program, and let $M$ and $N$ be positive (Hermitian) operators in $\mathcal{H}_{qvar(P)}$. \begin{enumerate}
\item If $tr(M\rho)\leq tr(N\llbracket P\rrbracket (\rho))$ for all $\rho\in\mathcal{D}(\mathcal{H}_{qvar(P)})$, then $M$ is called a precondition of $N$ with respect to $P$. \item $N$ is called the weakest precondition of $M$ with respect to $P$, written $N=wp.P.M$ if \begin{enumerate}\item $N$ is a precondition of $M$ with respect to $P$; and\item $N^\prime\sqsubseteq N$ whenever $N^\prime$ is a also precondition of $M$ with respect to $P$.  
\end{enumerate}\end{enumerate}
\end{defn}

$wp.P$ can be seen as the super-operator in $\mathcal{H}_{qvar(P)}$ defined as follows: for any positive operator $M$, $(wp.P)(M)=wp.P.M$ is given by clause 2) of the above definition, and $wp.P$ can be extended to the whole space $\mathcal{L}(\mathcal{H}_{qvar(P)})$ by linearity.  

The weakest precondition semantics of QGCL programs are given in the next proposition.

\begin{prop}\label{wpc}For any QGCL program $P$, and for any positive (Hermitian) operator $M$ in $\mathcal{H}_{qvar(P)}$, $wp.P.M$ is given as follows\begin{enumerate}\item $wp.\mathbf{abort}=0;$\item $wp.\mathbf{skip}=1;$\item $wp.(P_1;P_2)=wp.P_2;wp.P_1;$ \item $wp.U[\overline{q}]=U_{\overline{q}}^\dag\circ U_{\overline{q}};$\item $wp.(M[x\leftarrow\overline{q}]:\{P_m\})=\sum_m\left[wp.P_m;(M_m^\dag\circ M_m)\right];$\item $wp.(\square_{i=1}^n\ \overline{q}, |i\rangle\rightarrow P_i)\in \square_{i=1}^n\ |i\rangle\rightarrow wp.P_i.$
\end{enumerate}\end{prop}

Some cylindrical extensions of super-operators are used but unspecified in the above proposition because they can be recognised from the context. Again, \textquotedblleft$\in$" in the above clause 6) cannot be replaced by equality. 

\section{Quantum Choices: Superpositions of Programs}\label{QChoice}

\subsection{Definition and Example}

As explained in Sec.~\ref{Intro}, quantum choice may be defined based on quantum guarded command. 

\begin{defn}
Let $P$ and $P_i$ be programs for all $1\leq i\leq n$ such that $\overline{q}=qvar(P)$. If $\{|i\rangle\}$ is an orthonormal basis of $\mathcal{H}_{\overline{q}}$, and $\overline{q}\cap \bigcup_{i=1}^n qVar(P_i)=\emptyset$, then the quantum choice of $P_1,...,P_n$ according to $P$ along $\{|i\rangle\}$ is defined as $$\bigoplus_{i=1}^n\ P, |i\rangle \rightarrow P_i\stackrel{\triangle}{=}P; \square_{i=1}^n\ \overline{q}, |i\rangle \rightarrow P_i.$$ In particular, if $n=2$, then the quantum choice will be abbreviated to $P_1\ _{P[\overline{q}]}\oplus P_2$ or $P_1\ _{P}\oplus P_2$.
\end{defn}

\begin{exam}\label{q2w} Quantum walks have been extended to include multiple walkers and coins. These extended quantum walks can be conveniently written as QGCL programs with quantum choice. We consider two quantum walkers on a line sharing coins~\cite{XS12}. The Hilbert space of a single walker on a line is $\mathcal{H}=\mathcal{H}_p\otimes\mathcal{H}_c$, where $\mathcal{H}_p=span\{|x\rangle:x\in\mathbb{Z}\ (\rm integers)\}$ is the position space and $\mathcal{H}_c=span\{L,R\}$ is the coin space. Its step operator is $W=(T_L\otimes |L\rangle\langle L|+T_R\otimes |R\rangle\langle R|)(I_{\mathcal{H}_p}\otimes H),$ where $I_{\mathcal{H}_p}$ is the identity operator in $\mathcal{H}_p$, $$H=\frac{1}{\sqrt{2}}\left(\begin{array}{cc}1 & 1\\ 1&-1\end{array}\right)$$ is the $2\times 2$ Hadamard matrix, and $T_L, T_R$ are left- and right-translation, respectively; that is, $T_L|x\rangle=|x-1\rangle$, $T_R|x\rangle=|x+1\rangle$ for every $x\in \mathbb{Z}$. Then the Hilbert space of two walkers is $\mathcal{H}\otimes\mathcal{H}$, and if the two walkers are independent, then the step operator is $W\otimes W$. A two-qubit unitary operator $U$ can be introduced to entangle the two coins and it can be thought as that the two walkers are sharing coins. A step of two walkers sharing coins can be written as a QGCL program as follows: 
$$U[c_1,c_2];(T_L[q_1] _{H[c_1]}\oplus T_R[q_1]);(T_L[q_2] _{H[c_2]}\oplus T_R[q_2])$$ where $q_1, q_2$ are the position variables and $c_1,c_2$ the coin variables of the two walkers, respectively.      
\end{exam}

\subsection{Local Quantum Variables}

A quantum choice is defined as a \textquotedblleft coin" program followed by a quantum guarded command. A natural question would be: is it possible to move the \textquotedblleft coin" program to the end of a guarded command? To answer this question positively, we need to extend the syntax of QGCL by introducing block command with local quantum variables.

\begin{defn}\label{loc-def} Let $P$ be a QGCL program, let $\overline{q}\subseteq qvar(P)$ be a sequence of quantum variables, and let $\rho$ be a density operator in $\mathcal{H}_{\overline{q}}$. Then \begin{enumerate}\item The block command defined by $P$ restricted to $\overline{q}=\rho$ is: $$\mathbf{begin\ local}\ \overline{q}:=\rho;P\ \mathbf{end}.$$\item The quantum variables of the block command are: 
$$qvar\left(\mathbf{begin\ local}\ \overline{q}:=\rho;P\ \mathbf{end}\right)=qvar(P)\setminus\overline{q}.$$ \item The purely quantum denotational semantics of the block command is give as follows: $$\left\llbracket \mathbf{begin\ local}\ \overline{q}:=\rho;P\ \mathbf{end}\right\rrbracket (\sigma)=tr_{\mathcal{H}_{\overline{q}}}(\llbracket P\rrbracket (\sigma\otimes\rho))$$ for any density operator $\sigma$ in $\mathcal{H}_{qvar(P)\setminus\overline{q}}$.
\end{enumerate}\end{defn}

The following theorem shows that the \textquotedblleft coin" in a quantum choice can be move to the end of the guarded command if encapsulation in a block with local variables is allowed. 
 
\begin{thm}\label{local} \begin{equation}\label{local1}\bigoplus_{i=1}^n U[\overline{q}],|i\rangle\rightarrow P_i\equiv (\square_{i=1}^n U^\dag_{\overline{q}}|i\rangle\rightarrow P_i);U[\overline{q}].\end{equation} More generally, we have:
\begin{equation}\label{local2}\begin{split}\bigoplus_{i=1}^n P,|i\rangle\rightarrow P_i\equiv\ &\mathbf{begin\ local}\ \overline{r}:=|\varphi_0\rangle;\\ 
&\square_{i,j} |\psi_{ij}\rangle\rightarrow Q_{ij};U[\overline{q},\overline{r}]\ \mathbf{end}\end{split}\end{equation} for some new quantum variables $\overline{r}$, state $|\varphi_0\rangle\in\mathcal{H}_{\overline{r}}$, orthonormal basis $\{|\psi_{ij}\rangle\}$ of $\mathcal{H}_{\overline{q}}\otimes\mathcal{H}_{\overline{r}}$, programs $Q_{ij}$, and unitary operator $U$ in $\mathcal{H}_{\overline{q}}\otimes\mathcal{H}_{\overline{r}}$, where $\overline{q}=qvar(P)$.\end{thm}

\subsection{Quantum implementation of probabilistic choices}

We now examine the relation between probabilistic choice and quantum choice. To this end, we first extend the syntax of QGCL by adding probabilistic choice. 

\begin{defn}\label{prob-def}Let $P_i$ be a QGCL program for each $1\leq i\leq n$, and let $\{p_i\}_{i=1}^n$ be a sub-probability distribution; that is, $p_i\geq 0$ for each $1\leq i\leq n$ and $\sum_{i=1}^np_i\leq 1$. Then \begin{enumerate}\item The probabilistic choice of $P_1,...,P_n$ according to $\{p_i\}_{i=1}^n$ is 
$$\sum_{i=1}^n P_i@p_i.$$ \item The quantum variables of the choice are: $$qvar\left(\sum_{i=1}^n P_i@p_i\right)=\bigcup_{i=1}^n qvar(P_i).$$ \item The purely quantum denotational semantics of the choice is: 
$$\left\llbracket \sum_{i=1}^n P_i@p_i\right\rrbracket =\sum_{i=1}^n p_i\cdot \llbracket P_i\rrbracket.$$\end{enumerate}
\end{defn}

\begin{exam} (Continuation of Example~\ref{gmeas}; Probabilistic mixture of measurements) It is often required in quantum cryptographic protocols like BB84 to randomly choose between the measurement $M^0$ on a qubit in the computational basis and the measurement $M^1$ in the basis $|\pm\rangle$. If we perform measurement $M^i$ on qubit $|\psi\rangle$ and discard the outcomes of measurement, then we get $\rho_i=M_0^i|\psi\rangle\langle\psi|M_0^i +M_1^i|\psi\rangle\langle\psi|M_1^i$ for $i=0,1$. 
We now consider the unitary matrix $$U=\left(\begin{array}{cc}\sqrt{p} & \sqrt{q}\\ \sqrt{q} & -\sqrt{p}\end{array}\right)$$ on a qubit, where $p,q\geq 0$ and $p+q=1$. Let \begin{equation*}\begin{split}
P\stackrel{\triangle}{=}\mathbf{begin\ local}\ q:=|0\rangle;\ & q:=U[q]; \\ &\square_{i=0,1}q,|i\rangle\rightarrow M_i[q_1]\ \mathbf{end}
\end{split}\end{equation*} where $q,q_1$ are qubit variables. Then for any $|\psi\rangle\in\mathcal{H}_{q_1}$ and $i,j\in\{0,1\}$, we have: \begin{equation*}|\psi_{ij}\rangle\stackrel{\triangle}{=}M_{ij}(|\psi\rangle U|0\rangle)
=\sqrt{\frac{p}{2}}M_i^0 |\psi\rangle|0\rangle+\sqrt{\frac{q}{2}} M_j^1|\psi\rangle |1\rangle,\end{equation*} \begin{equation*}\begin{split}
&\llbracket P\rrbracket (|\psi\rangle\langle\psi|)=tr_{\mathcal{H}_q}\left(\sum_{i,j=0,1}|\psi_{ij}\rangle\langle\psi_{ij}|\right)\\
&=\sum_{i,j=0,1}\left(\frac{p}{2}M_i^0|\psi\rangle\langle\psi|M_i^0+\frac{q}{2}M_j^1|\psi\rangle\langle\psi|M_j^1\right)\\ &=p\rho_0 +q\rho_1.
\end{split}\end{equation*} So, program $P$ can be seen as a probabilistic mixture of measurements $M^0$ and $M^1$.
\end{exam}

As shown by the following theorem, if the \textquotedblleft coin" variables are treated as local variables, then a quantum choice degenerates to a probabilistic choice. 

\begin{thm}\label{proim} Let $qvar(P)=\overline{q}$. Then we have:
$$\mathbf{begin\ local}\ \overline{q}:=\rho;\bigoplus_{i=1}^nP,|i\rangle\rightarrow P_i\ \mathbf{end}\equiv \sum_{i=1}^nP_i@p_i$$ where $p_i=\langle i|\llbracket P\rrbracket (\rho)|i\rangle$ for every $1\leq i\leq n$.
\end{thm}

Conversely, for any probability distribution $\{p_i\}_{i=1}^n$, we can find an $n\times n$ unitary operator $U$ such that $p_i=|U_{i0}|^2$ $(1\leq i\leq n)$. So, it follows immediately from the above theorem that a probabilistic choice $\sum_{i=1}^nP_i@p_i$ can always be implemented by a quantum choice: $$\mathbf{begin\ local}\ \overline{q}:=|0\rangle;\bigoplus_{i=1}^n U[\overline{q}],|i\rangle\rightarrow P_i\ \mathbf{end}$$
where $\overline{q}$ is a family of new quantum variables with an $n-$dimensional state space. 

\section{Quantum Recursion}\label{QREC}

Now we need to further extend the syntax of QGCL. We first add a countable set of program names, ranged over by $X, Y, ...$, to the alphabet of QGCL, and then introduce the following:
\begin{defn} QGCL programs are defined by combining Definitions~\ref{syn-def}, \ref{loc-def}, \ref{prob-def} and the following two clauses: \begin{enumerate}\item Every program name $X$ is a program, and both $var(X)$ and $qvar(X)$ are given a priori.
\item If $P$ is a program and $X$ a program name such that $var(P)\subseteq var(X)$ and $qvar(P)\subseteq qvar(X)$, then $\mu X.P$ is a program, and $var(\mu X.P)=var(X),$ $qvar(\mu X.P)=qvar(X).$\end{enumerate}
\end{defn}

We consider a special case of quantum recursion, namely quantum loop, and show an interesting difference between quantum loops with classical control flows defined in~\cite{YF10} and quantum loops with quantum control flows. The quantum loops considered in~\cite{YF10} can be written as QGCL programs of the form: \begin{equation*}\begin{split}
Loop&=\mathbf{while}\ M[\overline{q}]=1\ \mathbf{do}\ \overline{q}:=U\overline{q}\\
&\stackrel{\triangle}{=}\mu X.M[x\leftarrow\overline{q}]:\{P_0=\mathbf{skip}, P_1=\overline{q}:=U\overline{q};X\}
\end{split}\end{equation*}
where $\overline{q}$ is a sequence of quantum variables, $M=\{M_0,M_1\}$ a binary (\textquotedblleft yes-no") measurement in $\mathcal{H}_{\overline{q}}$ and $U$ a unitary operator in $\mathcal{H}_{\overline{q}}$. The control flow of $Loop$ is determined by measurement $M$ in the loop guard: if the outcome of measurement is $0$ then $P_0$ is executed - the loop terminates; if the outcome of measurement is $1$ then $P_1$ is executed - the program executes the loop body $\overline{q}:=U\overline{q}$ and then runs into the loop again. Program $Loop$ can be approximated by a series of iterations $\{Q_n\}_{n=0}^\infty$ defined as follows:
\begin{equation}\label{iter}\begin{cases}
Q_0&\stackrel{\triangle}{=}\mathbf{abort},\\ Q_{n+1}&\stackrel{\triangle}{=}M[x\leftarrow\overline{q}]:\{P_0=\mathbf{skip},\\ &\ \ \ \ \ \ \ \ \ \ P_1^{(n+1)}=\overline{q}:=U\overline{q};Q_n\} \ (n\geq 0).
\end{cases}\end{equation} If the classical control flows of $Q_n$ $(n\geq 0)$ determined by the outcomes of measurement $M$ are replaced by quantum control flows defineed by quantum choices, then we obtain the following quantum iterations:    
\begin{equation*}\begin{cases}
Q^\prime_0&\stackrel{\triangle}{=}\mathbf{abort},\\ Q^\prime_{n+1}&\stackrel{\triangle}{=}\mathbf{skip} _{C[q_{n+1}]}\oplus (\overline{q}:=U\overline{q};Q^\prime_n)\ (n\geq 0)
\end{cases}\end{equation*}
where $C$ is a \textquotedblleft coin" $2\times 2$ unitary matrix. It is worth noting that we have to introduce a sequence $q_1,q_2,...$ of new qubit variables in order to well-define the quantum choices used in $Q_n^\prime$ $(n\geq 1)$. For each $n\geq 0$, since $var(Q_n^\prime)=\emptyset$ and $qvar(Q_n^\prime)=\overline{q}\cup\{q_1,...,q_n\}$, the semi-classical semantics $\lceil Q^\prime_n\rceil$ of $Q_n^\prime$ is an operator-valued function in $\mathcal{H}_{\overline{q}}\otimes\bigotimes_{i=1}^n \mathcal{H}_{q_i}$ over $\{\epsilon\}$. Suppose that the input is a state $|\psi\rangle$ in $\mathcal{H}_{\overline{q}}$, and all the auxiliary qubit variables $q_1,...,q_n$ are initialised in state $|0\rangle$. For simplicity of calculation, we take $C=H$ (the $2\times 2$ Hadamard matrix; see Example \ref{q2w}). Then \begin{equation*}\lceil Q_n^\prime\rceil (\epsilon)|\psi\rangle|0\rangle^n=\sum_{i=0}^{n-1}\frac{1}{\sqrt{2^{i+1}}}U^i|\psi\rangle|0\rangle^{n-i}|1\rangle^i.
\end{equation*} It is clear that we cannot directly define the semantics of a quantum loop as the limit of $\{Q_n^\prime\}_{n=0}^\infty$ because the state spaces of $Q_n^\prime$ are different for different $n$. To overcome this difficulty, a natural idea is to localise qubit variables $q_1,...,q_n$:   
\begin{equation*}Q^{\prime\prime}_n= \mathbf{begin\ local}\ q_1,...,q_n:= |0\rangle^n;Q^\prime_n\ \mathbf{end}.\end{equation*} But such a localisation makes the quantum iterations degenerate to probabilistic iterations:
\begin{equation*}\llbracket Q_n^{\prime\prime}\rrbracket (\rho)= \sum_{i=0}^{n-1}\frac{1}{2^{i+1}}U^i\rho (U^\dag)^i.\end{equation*}
This gives an example showing that quantum loops, or more generally quantum recursions, with quantum control flows are much harder to deal with than those with classical control flows. Due to the limited space, a more detailed treatment of quantum recursion is postponed to another paper. 
\section{Conclusions}

Three new quantum program constructs - quantum guarded command, quantum choice and quantum recursion - are defined in this paper. We believe that introducing these constructs is a significant step toward the full realisation of \textquotedblleft quantum control" in quantum programming. In the further studies, we will consider quantum recursions with quantum controls in detail, and we will establish various algebraic laws for QGCL programs that can be used in program transformations and compilation. A quantum Floyd-Hoare logic was built in~\cite{Y11} for quantum programs with only classical control flows. So, another interesting topic for further studies would be to extend this logic so that it can also be applied to programs with quantum control flows.

\newpage

\section*{Appendix: Proofs}

\subsection{Proof of Lemma~\ref{gulem}} 

We start with an auxiliary equality. Put: $$\overline{F}\stackrel{\triangle}{=}\sum_{\delta_1\in\Delta_1,...,\delta_n\in\Delta_n}F(\oplus_{i=1}^n \delta_i)^\dag\cdot F(\oplus_{i=1}^n\delta_i).$$ For any $|\varphi\rangle, |\psi\rangle\in\mathcal{H}\otimes\mathcal{H}_s$, we can write: \begin{equation*}\begin{split}|\varphi\rangle&=\sum_{i=1}^n|\varphi_i\rangle|i\rangle,\\ |\psi\rangle&=\sum_{i=1}^n|\psi_i\rangle|i\rangle\end{split}\end{equation*} where $|\varphi_i\rangle, |\psi_i\rangle\in\mathcal{H}$ for each $1\leq i\leq n$. Then \begin{equation}\label{gud-pr}\begin{split}&\langle\varphi|\overline{F}|\psi\rangle=\sum_{\delta_1,...,\delta_n}\langle\varphi| F(\oplus_{i=1}^n\delta_i)^\dag\cdot F(\oplus_{i=1}^n\delta_i)|\psi\rangle\\
&=\sum_{\delta_1,...,\delta_n}\sum_{i,i^\prime =1}^n\left(\prod_{k\neq i}\lambda^\ast_{k\delta_k}\right)\left(\prod_{k\neq i^\prime}\lambda_{k\delta_k}\right) \\ &\ \ \ \ \ \ \ \ \ \ \ \ \ \ \ \ \ \ \ \ \ \ \ \ \ \ \ \ \ \ \langle\varphi_i|
F_{i}(\delta_i)^\dag F_{i^\prime}(\delta_{i^\prime})|\psi_{i^\prime}\rangle\langle i|i^\prime\rangle\\ &=\sum_{\delta_1,...,\delta_n}\sum_{i=1}^n\left(\prod_{k\neq i}|\lambda_{k\delta_k}|^2\right) \langle\varphi_i|F_{i}(\delta_i)^\dag F_{i}(\delta_{i})|\psi_{i}\rangle\\
&=\sum_{i=1}^n\sum_{\delta_1,...,\delta_{i-1},\delta_{i+1},...,\delta_n}\left(\prod_{k\neq i}|\lambda_{k\delta_k}|^2\right)\\ &\ \ \ \ \ \ \ \ \ \ \ \ \ \ \ \ \ \ \ \ \ \ \ \ \ \ \ \ \ \ \sum_{\delta_i} \langle\varphi_i|F_{i}(\delta_i)^\dag F_{i}(\delta_{i})|\psi_{i}\rangle\\
&=\sum_{i=1}^n \sum_{\delta_i} \langle\varphi_i|F_{i}(\delta_i)^\dag F_{i}(\delta_{i})|\psi_{i}\rangle\\
&=\sum_{i=1}^n\langle\varphi_i|\sum_{\delta_i}F_{i}(\delta_i)^\dag F_{i}(\delta_{i})|\psi_{i}\rangle
\end{split}\end{equation} because for each $k$, we have: $$\sum_{\delta_k}|\lambda_{k\delta_k}|^2=1,$$ and thus \begin{equation}\label{coeff}\sum_{\delta_1,...,\delta_{i-1},\delta_{i+1},...,\delta_n}\left(\prod_{k\neq i}|\lambda_{k\delta_k}|^2\right)
=\prod_{k\neq i}\left(\sum_{\delta_k}|\lambda_{k\delta_k}|^2\right)=1.\end{equation}

(1) We now prove that $F$ is a semi-classical semantic function in $\mathcal{H}\otimes\mathcal{H}_s$ over $\bigoplus_{i=1}^n\Delta_n$. It suffices to show that $\overline{F}\sqsubseteq I_{\mathcal{H}\otimes\mathcal{H}_s}$; that is, $\langle\varphi|\overline{F}|\varphi\rangle\leq\langle\varphi|\varphi\rangle$ for each $|\varphi\rangle\in\mathcal{H}\otimes\mathcal{H}_s$. In fact, for each $1\leq i\leq n$, since $F_i$ is a semi-classical semantic function, we have: $$\sum_{\delta_i}F_{i}(\delta_i)^\dag F_{i}(\delta_i)\sqsubseteq I_\mathcal{H},$$ $$\langle\varphi_i|\sum_{\delta_i}F_{i}(\delta_i)^\dag F_{i}(\delta_i)|\varphi_i\rangle\leq\langle\varphi_i|\varphi_i\rangle.$$ Then it follows from Eq.~(\ref{gud-pr}) that 
\begin{equation*}\langle\varphi|\overline{F}|\varphi\rangle\leq\sum_{i=1}^n\langle\varphi_i|\varphi_{i}\rangle=\langle\varphi|\varphi\rangle.\end{equation*}

(2) For the case where all $F_i$ $(1\leq i\leq n)$ are full, we have: $$\sum_{\delta_i}F_{i}(\delta_i)^\dag F_{i}(\delta_i)=I_\mathcal{H}$$ for all $1\leq i\leq n$, and it follows from Eq.~(\ref{gud-pr}) that $$\langle\varphi|\overline{F}|\psi\rangle=\sum_{i=1}^n\langle\varphi_i|\psi_i\rangle=\langle\varphi|\psi\rangle.$$ So, it holds that $\overline{F}=I_{\mathcal{H}\otimes\mathcal{H}_s}$ by arbitrariness of $|\varphi\rangle$ and $\psi\rangle$, and $F$ is full. 

\subsection{Proof of Proposition~\ref{qsem}}

Clauses 1) - 4) are obvious.

5) By definition, for any partial density operator $\rho$ in $\mathcal{H}_{qvar(P)}$, we have: \begin{equation*}\begin{split}&\llbracket M[x\overline{q}]:\{P_m\}
\rrbracket (\rho)\\ &=\sum_m\sum_{\delta\in\Delta(P_m)}\lceil P\rceil (\delta[x\leftarrow m])\rho\lceil P\rceil (\delta[x\leftarrow m])^\dag\\
&=\sum_m\sum_{\delta\in\Delta(P_m)}(\lceil P_m\rceil (\delta)\otimes I_{qvar(P)\setminus qvar(P_m)})\\ &\ \ \ \ \ \ \ \ \ \ \ \ \ \ \ \ (M_m\otimes I_{qvar(P)\setminus\overline{q}})\rho (M_m^\dag\otimes I_{qvar(P)\setminus\overline{q}})\\ &\ \ \ \ \ \ \ \ \ \ \ \ \ \ \ \ \ \ \ \ \ \ \ \ \ \ \ \ \ \ \ \ \ \ \ \ \ 
(\lceil P_m\rceil (\delta)^\dag\otimes I_{V\setminus qvar(P_m)})\\ &=\sum_m\sum_{\delta\in\Delta(P_m)}(\lceil P_m\rceil (\delta)\otimes I_{qvar(P)\setminus qvar(P_m)})\\ &\ \ \ \ \ \ \ \ \ \ \ \ \ \ \ \ \ \ \ \ (M_m\rho M_m^\dag)
(\lceil P_m\rceil (\delta)^\dag\otimes I_{V\setminus qvar(P_m)})\\ &=\sum_m\llbracket P_m\rrbracket (M_m\rho M_m^\dag)\\ &=\left(\sum_m(M_m\circ M_m^\dag);\llbracket P_m\rrbracket\right)(\rho).
\end{split}\end{equation*}

6) For simplicity of the presentation, we write: $$P\stackrel{\triangle}{=}\square_{i=1}^n\overline{q}, |i\rangle\rightarrow P_i.$$ By definition, we obtain: $$\llbracket P\rrbracket= \mathcal{E}(\lceil \square_{i=1}^n\overline{q}, |i\rangle\rightarrow P_i\rceil).$$ Since $$\lceil \square_{i=1}^n\overline{q}, |i\rangle\rightarrow P_i\rceil=\square_{i=1}^n\overline{q}, |i\rangle\rightarrow \lceil P_i\rceil$$ and $\lceil P_i\rceil\in\mathbb{F}(\llbracket P_i\rrbracket)$ for every $1\leq i\leq n$, it holds that 
\begin{equation*}\begin{split}\llbracket P\rrbracket\in\{\mathcal{E}(&\square_{i=1}^n\ |i\rangle\rightarrow F_i):F_i\in\mathbb{F}(\llbracket P_i\rrbracket)\\ &{\rm for\ every}\ 1\leq i\leq n\}=\square_{i=1}^n\ |i\rangle\rightarrow \llbracket P_i\rrbracket.\end{split}
\end{equation*}

\subsection{Proof of Proposition~\ref{wpc}}

The proof is based on the following key lemma by D'Hondt and Panangaden~\cite{DP06}.

\begin{lem}\label{DPL} If the semantic function $\llbracket P\rrbracket$ of program $P$ has the Kraus operator-sum representation: $\llbracket P\rrbracket =\sum_j E_j\circ E_j^\dag,$ then we have: $wp.P=\sum_jE_j^\dag\circ E_j.$\end{lem}

Now we start to prove Proposition~\ref{wpc}. Clauses 1) - 4) are immediate from Proposition~\ref{qsem} and Lemma~\ref{DPL}.

5) Suppose that for every $m$, $$\llbracket P_m\rrbracket =\sum_m E_{mi_m}\circ E_{mi_m}^\dag.$$ Then by Proposition~\ref{qsem} 5) we have: \begin{equation*}\begin{split}& 
\llbracket M[x\leftarrow\overline{q}]:\{P_m\}\rrbracket =\sum_m\left[(M_m\circ M_m^\dag);\llbracket P_m\rrbracket\right]\\ 
&=\sum_m \left[(M_m\circ M_m^\dag);\sum_{i_m}\left(E_{mi_m}\circ E_{mi_m}^\dag\right)\right]\\
&=\sum_m\sum_{i_m}\left(E_{mi_m}M_m\right)\circ \left(M_m^\dag E_{mi_m}^\dag\right)\\
&=\sum_m\sum_{i_m}\left(E_{mi_m}M_m\right)\circ \left(E_{mi_m}M_m\right)^\dag.
\end{split}\end{equation*} Using Lemma~\ref{DPL} we obtain: 
\begin{equation*}\begin{split}&wp.(M[x\leftarrow\overline{q}];\{P_m\})=\sum_m\sum_{i_m}\left(E_{mi_m}M_m\right)^\dag\circ\left(E_{mi_m}M_m\right)\\
&=\sum_m\sum_{i_m}\left(M_m^\dag E_{mi_m}^\dag\right)\circ\left(E_{mi_m}M_m\right)\\
&=\sum_m\left[\sum_{i_m}\left(E_{mi_m}^\dag\circ E_{mi_m}\right); \left(M_{m}^\dag\circ M_m\right)\right]\\
&=\sum_m \left[wp.P_m;(M_m^\dag\circ M_m)\right].
\end{split}\end{equation*}

6) For each $1\leq i\leq n$, assume that the semi-classical semantics of $P_i$ is the function $\lceil P_i\rceil$ over $\Delta=\{j_i\}$ such that $$\lceil P_i\rceil (j_i)=E_{ij_i}$$ for every $j_i$. Then 
by Definition~\ref{den-def} we obtain: 
$$\llbracket P_i\rrbracket =\sum_{j_i}E_{ij_i}\circ E_{ij_i}^\dag,$$ and it follows from Lemma~\ref{DPL} that $$wp.P_i=\sum_{j_i}E_{ij_i}^\dag\circ E_{ij_i}.$$ 

For any $|\varphi\rangle=\sum_{i=1}^n|\varphi_i\rangle|i\rangle,$ where $|\varphi_i\rangle\in\mathcal{H}_{qvar(P_i)}$ $(1\leq i\leq n)$, we define:   
\begin{equation*}\begin{split}G_{j_1...j_n}(|\varphi\rangle)&=\sum_{i=1}^n\zeta_i E_{ij_i}^\dag |\varphi_i\rangle|i\rangle,\\
\zeta_i&=\prod_{k\neq i}\delta_{kj_k},\end{split}\end{equation*}
\begin{equation}\label{coef3}\begin{split}\delta_{kj_k}&=\sqrt{\frac{tr (E_{kj_k}^\dag)^\dag E_{kj_k}^\dag}{\sum_{l_k}(E_{kl_k}^\dag)^\dag E_{kl_k}^\dag}}=\sqrt{\frac{tr E_{kj_k}^\dag E_{kj_k}}{\sum_{l_k} E_{kl_k}^\dag E_{kl_k}}}=\lambda_{kj_k}
\end{split}\end{equation} and $\lambda_{kj_k}$'s are defined by Eq.~(\ref{coef1}).
By Definitions~\ref{qgu} and \ref{def-gsup} we have: $$\sum_{j_1,...,j_n}G_{j_1...j_n}\circ G_{j_1...j_n}^\dag\in\square_{i=1}^n\ |i\rangle\rightarrow wp.P_i.$$ On the other hand, by Definitions \ref{defsem} (5) and~\ref{den-def} we have: 
$$\llbracket\square_{i=1}^n\ q,|i\rangle\rightarrow P_i\rrbracket=\sum_{j_1,...,j_n}F_{j_1...j_n}\circ F_{j_1...j_n}^\dag$$ where $F_{j_1...j_n}$'s are defined by Eq.~(\ref{coef0}). Applying Lemma~\ref{DPL} once again, we obtain: 
$$wp.(\square_{i=1}^n\ q,|i\rangle\rightarrow P_i)=\sum_{j_1,...,j_n}F_{j_1...j_n}^\dag\circ F_{j_1...j_n}.$$ So, we now only need to prove that $$G_{j_1...j_n}=F_{j_1...j_n}^\dag$$ for all $j_1,...,j_n$. In fact, for any $|\psi\rangle=\sum_{i=1}^n |\psi_i\rangle|i\rangle$ with $|\psi_i\rangle\in\mathcal{H}_{qvar(P_i)}$ $(1\leq i\leq n)$, it holds that 
\begin{equation*}\begin{split}(G_{j_1...j_n}|\varphi\rangle,|\psi\rangle)&=(\sum_{i=1}^n\zeta_i E_{ij_i}^\dag |\varphi_i\rangle|i\rangle,\sum_{i=1}^n|\psi_i\rangle|i\rangle)\\ 
&=\sum_{i,i^\prime}\zeta_i^\ast(E_{ij_i}^\dag |\varphi_i\rangle,|\psi_{i^\prime}\rangle)\langle i|i^\prime\rangle\\
&=\sum_i\zeta_i(E_{ij_i}^\dag |\varphi_i\rangle,|\psi_{i}\rangle)\\
&=\sum_i\zeta_i(|\varphi_i\rangle,E_{ij_i}|\psi_{i}\rangle)\\
&=\sum_{i,i^\prime}\zeta_i(|\varphi_i\rangle,E_{i^\prime j_{i^\prime}}|\psi_{i^\prime}\rangle)\langle i|i^\prime\rangle\\
&=(\sum_{i=1}^n |\varphi_i\rangle|i\rangle,\sum_{i=1}^n \zeta_i E_{ij_i}|\psi_i\rangle|i\rangle)\\ 
&=(|\varphi\rangle,F_{j_1...j_n}|\psi\rangle)
\end{split}\end{equation*} because $\zeta_i$'s are real numbers, and it follows from Eq.~(\ref{coef3}) that $$\zeta_i=\prod_{k\neq i}\lambda_{kj_k}$$ for each $1\leq i\leq n$. Thus, we complete the proof. 

\subsection{Proof of Theorem~\ref{local}}

We first prove Eq.~(\ref{local1}). Assume that $\lceil P_i\rceil$ is the operator-valued function over $\Delta_i$ such that $\lceil P_i\rceil (\delta_i)=F_{i\delta_i}$ for each $\delta_i\in\Delta_i$ $(1\leq i\leq n)$. We write: $$P=\square_{i=1}^n U_{\overline{q}}^\dag |i\rangle\rightarrow P_i.$$ Then for any $|\psi\rangle=\sum_{i=1}^n |\psi_i\rangle|i\rangle$, where $|\psi_i\rangle\in\mathcal{H}_V$ $(1\leq i\leq n)$, and $V=\bigcup_{i=1}^n qvar(P_i)$, we have: \begin{equation*}\begin{split}
&\lceil P\rceil(\oplus_{i=1}^n\delta_i)|\psi\rangle
\\ &=\lceil P\rceil(\oplus_{i=1}^n\delta_i)\left[\sum_{i=1}^n|\psi_i\rangle\left(\sum_{j=1}^n U_{ij}(U_{\overline{q}}^\dag |j\rangle)\right)\right]
\\ &=\lceil P\rceil(\oplus_{i=1}^n\delta_i)\left[\sum_{j=1}^n\left(\sum_{i=1}^n U_{ij}|\psi_i\rangle\right)(U_{\overline{q}}^\dag |j\rangle)\right]
\\ &=\sum_{j=1}^n \left(\prod_{k\neq j}\lambda_{k\delta_k}\right)F_{j\delta_j}\left(\sum_{i=1}^n U_{ij}|\psi_i\rangle\right)(U_{\overline{q}}^\dag |j\rangle).
\end{split}\end{equation*} Let $LHS$ and $RHS$ stand for the left and right hand side of Eq.~(\ref{local1}), respectively. Then it holds that \begin{equation*}\begin{split}
&\lceil RHS\rceil (\oplus_{i=1}^n\delta_i)|\psi\rangle= U_{\overline{q}}(\lceil P\rceil (\oplus_{i=1}^n\delta_i)|\psi\rangle)\\
& =\sum_{j=1}^n \left(\prod_{k\neq j}\lambda_{k\delta_k}\right)F_{j\delta_j}\left(\sum_{i=1}^n U_{ij}|\psi_i\rangle\right)|j\rangle\\
&=\lceil\square_{i=1}^n |i\rangle\rightarrow P_i\rceil (\oplus_{i=1}^n\delta_i)\left[\sum_{j=1}^n\left(\sum_{i=1}^n U_{ij}|\psi_i\rangle\right)|j\rangle\right]\\
&=\lceil\square_{i=1}^n |i\rangle\rightarrow P_i\rceil (\oplus_{i=1}^n\delta_i)\left[\sum_{i=1}^n |\psi_i\rangle\sum_{j=1}^n\left(U_{ij}|j\rangle\right)\right]\\
&=\lceil\square_{i=1}^n |i\rangle\rightarrow P_i\rceil (\oplus_{i=1}^n\delta_i)\left(\sum_{i=1}^n |\psi_i\rangle(U_{\overline{q}}|i\rangle)\right)\\
&=\lceil LHS\rceil (\oplus_{i=1}^n\delta_i) |\psi\rangle.
\end{split}\end{equation*} Consequently, it follows that $\llbracket LHS\rrbracket =\llbracket RHS\rrbracket$ and we complete the proof of Eq.~(\ref{local1}).

Now we are ready to prove Eq.~(\ref{local2}). Since $\llbracket P\rrbracket$ is a super-operator in $\mathcal{H}_{\overline{q}}$, there must be a family of quantum variables $\overline{r}$, a pure state $|\varphi_0\rangle\in\mathcal{H}_{\overline{r}}$, a unitary operator $U$ in $\mathcal{H}_{\overline{q}}\otimes\mathcal{H}_{\overline{r}}$, and a projection operator $K$ onto some closed subspace $\mathcal{K}$ of $\mathcal{H}_{\overline{r}}$ such that $$\llbracket P\rrbracket (\rho)=tr_{\mathcal{H}_{\overline{r}}}(KU(\rho\otimes |\varphi_0\rangle\langle\varphi_0|)U^\dag K)$$ for all density operators $\rho$ in $\mathcal{H}_{\overline{q}}$ (see the system-environment model of super-operators, Eq.~(8.38) in \cite{NC00}). We choose an orthonormal basis of $\mathcal{K}$ and then extend it to an orthonormal basis $\{|j\rangle\}$ of $\mathcal{H}_{\overline{r}}$. Define pure states $|\psi_{ij}\rangle=U^\dag |ij\rangle$ for all $i, j$ and programs $$Q_{ij}=\begin{cases}P_i\ &{\rm if}\ |j\rangle\in\mathcal{K},\\ \mathbf{abort}\ &{\rm if}\ |j\rangle\notin \mathcal{K}.\end{cases}$$ Then by a routine calculation we have:  
\begin{equation}\label{local3}\llbracket\square_{i,j}|ij\rangle\rightarrow Q_{ij}\rrbracket (\sigma)=\llbracket \square_i|i\rangle\rightarrow P_i\rrbracket (K\sigma K)
\end{equation} for any $\sigma\in\mathcal{H}_{\overline{q}\cup\overline{r}\cup V}$, where $V=\bigcup_{i=1}^n qvar(P_i)$. We now write $RHS$ for the right hand side of Eq.~(\ref{local2}). Combining Eqs~(\ref{local1}) and (\ref{local3}), we obtain: \begin{equation*}\begin{split}\llbracket &RHS\rrbracket (\rho)\\ &=tr_{\mathcal{H}_{\overline{r}}}\left(\llbracket \square_{i,j}U^\dag |ij\rangle\rightarrow Q_{ij};U[\overline{q},\overline{r}]\rrbracket (\rho\otimes |\varphi_0\rangle\langle\varphi_0|\right)\\ & =tr_{\mathcal{H}_{\overline{r}}}\left(\llbracket \bigoplus_{i,j}U^\dag U[\overline{q},\overline{r}], |ij\rangle\rightarrow Q_{ij}\rrbracket (\rho\otimes |\varphi_0\rangle\langle\varphi_0|)\right)\\ & =tr_{\mathcal{H}_{\overline{r}}}\left(\llbracket \square_{i,j} |ij\rangle\rightarrow Q_{ij}\rrbracket (U(\rho\otimes |\varphi_0\rangle\langle\varphi_0|)U^\dag)\right)\\ 
& =tr_{\mathcal{H}_{\overline{r}}}\llbracket \square_{i} |i\rangle\rightarrow P_{i}\rrbracket (KU(\rho\otimes |\varphi_0\rangle\langle\varphi_0|)U^\dag K)\\ 
& =\llbracket \square_{i} |i\rangle\rightarrow P_{i}\rrbracket (tr_{\mathcal{H}_{\overline{r}}}(KU(\rho\otimes |\varphi_0\rangle\langle\varphi_0|)U^\dag K))\\ 
& =\llbracket \square_{i} |i\rangle\rightarrow P_{i}\rrbracket (\llbracket P\rrbracket (\rho))\\ & =\llbracket \bigoplus_{i} P, |i\rangle\rightarrow P_{i}\rrbracket (\rho)
\end{split}\end{equation*} for all density operators $\rho$ in $\mathcal{H}_{\overline{q}}$. Therefore, Eq.~(\ref{local2}) is proved.

\subsection{Proof of Theorem~\ref{proim}}

To simplify the presentation, we write: $$R\stackrel{\triangle}{=}\square_{i=1}^n\ \overline{q},|i\rangle\rightarrow P_i,$$ and assume that $\lceil P_i\rceil$ is the operator-valued function over $\Delta_i$ such that $\lceil P_i\rceil(\delta_i)=E_{i\delta_i}$ for each $\delta_i\in\Delta_i$. Let $|\psi\rangle\in\mathcal{H}_{\bigcup_{i=1}^n qvar(P_i)}$ and $|\varphi\rangle\in\mathcal{H}_{\overline{q}}$. We can write:  
$$|\varphi\rangle=\sum_{i=1}^n\alpha_i |i\rangle$$ for some complex numbers $\alpha_i$ $(1\leq i\leq n)$. Then for any $\delta_i\in\Delta_i$ $(1\leq i\leq n)$, we have: \begin{equation*}\begin{split}
|\Psi_{\delta_1...\delta_n}\rangle&\stackrel{\triangle}{=}\lceil R\rceil (\oplus_{i=1}^n\delta_i)(|\psi\varphi\rangle)\\ &=\lceil R\rceil (\oplus_{i=1}^n\delta_i)\left(\sum_{i=1}^n\alpha_i |\psi i\rangle\right)\\
&=\sum_{i=1}^n\alpha_i\left(\prod_{k\neq i}\lambda_{k\delta_k}\right)E_{i\sigma_i}|\psi\rangle |i\rangle
\end{split}\end{equation*} where $\lambda_{i\delta_i}$'s are defined as in Eq.~(\ref{coef1}), 
\begin{equation*}\begin{split}
|\Psi_{\delta_1...\delta_n}\rangle\langle\Psi_{\delta_1...\delta_n}|=\sum_{i,j=1}^n &\alpha_i\alpha_j^\ast\left(\prod_{k\neq i}\lambda_{k\delta_k}\right)\left(\prod_{k\neq j}\lambda_{k\delta_k}\right)\\ &\ \ \ \ \ \ \ \ \ E_{i\delta_i}|\psi\rangle\langle\psi|E_{j\delta_j}^\dag\otimes |i\rangle\langle j|,\end{split}\end{equation*} and it follows that 
\begin{equation*}\begin{split}
&tr_{\mathcal{H}_{\overline{q}}}|\Psi_{\delta_1...\delta_n}\rangle\langle\Psi_{\delta_1...\delta_n}|\\ &=\sum_{i=1}^n|\alpha_i|^2 \left(\prod_{k\neq i}\lambda_{k\delta_k}\right)^2 E_{i\delta_i}|\psi\rangle\langle\psi|E_{i\delta_i}^\dag.\end{split}\end{equation*} Using Eq.~(\ref{coeff}), we obtain:
\begin{equation}\label{trr}\begin{split}
&tr_{\mathcal{H}_{\overline{q}}}\llbracket R\rrbracket (|\psi\varphi\rangle\langle\varphi\psi|)\\ &=tr_{\mathcal{H}_{\overline{q}}}\left(\sum_{\delta_1,...,\delta_n}|\Psi_{\delta_1...\delta_n}\rangle\langle\Psi_{\delta_1...\delta_n}|\right)\\ &=\sum_{\delta_1,...,\delta_n}tr_{\mathcal{H}_{\overline{q}}}|\Psi_{\delta_1...\delta_n}\rangle\langle\Psi_{\delta_1...\delta_n}|\\ 
&=\sum_{i=1}^n|\alpha_i|^2\left[\sum_{\delta_1,...,\delta_{i-1},\delta_{i+1},...,\delta_n}\left(\prod_{k\neq i}\lambda_{k\delta_k}\right)^2\right]\\ &\ \ \ \ \ \ \ \ \ \ \ \ \ \ \ \ \ \ \ \ \ \ \ \ \ \ \ \ \ \ \cdot \left[\sum_{\delta_i}E_{i\delta_i}|\psi\rangle\langle\psi|E_{i\delta_i}^\dag\right]\\ &=\sum_{i=1}^n|\alpha_i|^2\llbracket P_i\rrbracket (|\psi\rangle\langle\psi|).\end{split}
\end{equation}

Now we do spectral decomposition for  $\llbracket P\rrbracket (\rho)$ and assume that $$\llbracket P\rrbracket (\rho)=\sum_l s_l|\varphi_l\rangle\langle\varphi_l|.$$ We further write: $$|\varphi_l\rangle=\sum_i\alpha_{li}|i\rangle$$ for every $l$. For any density operator $\sigma$ in $\mathcal{H}_{\bigcup_{i=1}^n qvar(P_i)}$, we can write $\sigma$ in the form of $$\sigma=\sum_m r_m|\psi_m\rangle\langle\psi_m|.$$ Then using Eq.~(\ref{trr}), we get: \begin{equation*}\begin{split}
&\llbracket\mathbf{begin\ local}\ \overline{q}:=\rho;\square_{i=1}^n P,|i\rangle\rightarrow P_i\ \mathbf{end}\rrbracket(\sigma)\\ &=tr_{\mathcal{H}_{\overline{q}}}\llbracket P;R\rrbracket (\sigma\otimes\rho)\\
&=tr_{\mathcal{H}_{\overline{q}}}\llbracket R\rrbracket(\sigma\otimes\llbracket P\rrbracket(\rho))\\
&=tr_{\mathcal{H}_{\overline{q}}}\llbracket R\rrbracket\left(\sum_{m,l}r_ms_l|\psi_m\varphi_l\rangle\langle\varphi_l\psi_m|\right)\\
&=\sum_{m,l}r_ms_ltr_{\mathcal{H}_{\overline{q}}}\llbracket R\rrbracket (|\psi_m\varphi_l\rangle\langle\varphi_l\psi_m|)\\ 
&=\sum_{m,l}r_ms_l\sum_{i=1}^n|\alpha_{li}|^2\llbracket P_i\rrbracket(|\psi_m\rangle\langle\psi_m|)\\
&=\sum_l\sum_{i=1}^ns_l|\alpha_{li}|^2\llbracket P_i\rrbracket\left(\sum_m r_m|\psi_m\rangle\langle\psi_m|\right)\\
&=\sum_l\sum_{i=1}^ns_l|\alpha_{li}|^2\llbracket P_i\rrbracket(\sigma)\\
&=\sum_{i=1}^n\left(\sum_ls_l|\alpha_{li}|^2\right)\llbracket P_i\rrbracket(\sigma)\\
&=\left\llbracket\sum_{i=1}^n P_i@p_i\right\rrbracket(\sigma),
\end{split}\end{equation*} where \begin{equation*}\begin{split}
p_i&=\sum_{l}s_l|\alpha_{li}|^2
=\sum_ls_l\langle i|\varphi_l\rangle\langle\varphi_l|i\rangle\\ &=\langle i|\left(\sum_ls_l|\varphi_l\rangle\langle\varphi_l|\right)|i\rangle=\langle i|\llbracket P\rrbracket (\rho)|i\rangle.\end{split}\end{equation*}
\end{document}